\documentclass[a4paper,fleqn,usenatbib]{mnras}

\usepackage[T1]{fontenc}
\usepackage{ae,aecompl}

\usepackage{graphicx}	% Including figure files
\usepackage{amsmath}	% Advanced maths commands
\usepackage{amssymb}	% Extra maths symbols

\title[Rayleigh Scattered Ly$\alpha$ in AGN]{Polarization of Rayleigh Scattered Ly$\alpha$ 
in Active Galactic Nuclei}

\author[S.-J. Chang et al.]{
Seok-Jun Chang,$^{1}$\thanks{E-mail: csj607@gmail.com}
Hee-Won Lee$^{1}$\thanks{E-mail: hwlee@sejong.ac.kr}
and Yujin Yang$^{2,3}$
\\
$^{1}$ Department of Physics and Astronomy, Sejong University, Seoul, Korea\\
$^{2}$ Korea Astronomy and Space Science Institute, Daejon, Korea\\
$^{3}$ Korea University of Science and Technology (UST), 217 Gajeong-ro Yuseong-gu, Daejeon 34113, Korea
}

\date{Accepted XXX. Received YYY; in original form ZZZ}

\pubyear{2016}

\begin{document}
\label{firstpage}
\pagerange{\pageref{firstpage}--\pageref{lastpage}}
\maketitle

% Abstract of the paper
\begin{abstract}
%Ly$\alpha$ is the most prominent emission line exhibited by active galactic nuclei (AGNs).
%Active galactic nuclei (AGNs) are divided into two types, in which 
%type 1 AGNs show both broad and narrow emission lines and 
%type 2 AGNs exhibit only narrow emission lines. 
The unification scheme of active galactic nuclei (AGNs) invokes an optically 
thick molecular torus component hiding the broad emission line region. 
Assuming the presence of 
a thick neutral component in the molecular torus characterized by
a \ion{H}{I} column density > $10^{22}{\rm\ cm^{-2}}$,
 we propose that
far UV radiation around Ly$\alpha$ can be
significantly polarized through Rayleigh scattering.  Adopting a Monte Carlo 
technique we compute polarization of Rayleigh 
scattered radiation near Ly$\alpha$ in a thick neutral region in the shape
of a slab and a cylindrical shell.
It is found that radiation near Ly$\alpha$ Rayleigh reflected 
from a very thick slab
can be significantly polarized in a fairly large range of wavelength 
$\Delta\lambda\sim 50$ \AA\ exhibiting a flux profile similar 
to the incident one.
Rayleigh transmitted radiation in a slab is characterized by
the central dip with a complicated polarization behavior.
The optically thick part near Ly$\alpha$ center is polarized 
in the direction perpendicular to the slab normal, which is in contrast to
weakly polarized wing parts in the direction parallel to the slab normal.
 A similar polarization flip phenomenon is also found in the case of 
a tall cylindrical shell, in which the spatial diffusion along the vertical direction
near the inner cylinder wall for core photons leads to a tendency
of the electric field aligned to the direction perpendicular to the vertical axis.
Observational implications are briefly discussed including 
spectropolarimetry of the quasar PG~1630+377 by Koratkar et al. in 1990
where Ly$\alpha$ is strongly polarized with no other emission lines polarized.
\end{abstract}

% Select between one and six entries from the list of approved keywords.
% Don't make up new ones.

%%%%%%%%%%%%%%%%%%%%%%%%%%%%%%%%%%%%%%%%%%%%%%%%%%

%%%%%%%%%%%%%%%%% BODY OF PAPER %%%%%%%%%%%%%%%%%%

\section{Introduction}

The spectra of active galactic nuclei (AGNs) are characterized by a nonthermal
continuum with prominent emission lines. A huge range of ionization is apparent
in the emission line spectra encompassing low ionization species such as 
\ion{Mg}{ii}$\lambda$2800 and high ionization lines like 
\ion{O}{vi}$\lambda$1034. AGNs are powered by the accretion
process around a supermassive black hole forming  a geometrically
thin but optically thick disk \citep{blandford90}. 
The emission lines are classified according 
to their widths where broad lines exhibit a typical width of $10^4\ {\rm km\ 
s^{-1}}$ and narrow emission lines show an order of magnitude smaller width 
of $500{\rm\ km\ s^{-1}}$.

AGNs are usually classified into two 
types according to the widths of emission lines. Type 1 AGNs show both broad 
and narrow emission lines whereas type 2 AGNs 
exhibit only narrow emission lines. According to unification models of AGNs 
all AGNs possess a highly thick molecular torus outside
the broad emission line region (e.g. \citealt{netzer15}). In these models, 
the two types of AGNs are 
interpreted as an orientation effect toward the observer's line of sight. 
Type I AGNs are those viewed by the observers with the line of sight near 
the polar direction whereas to the observers near the equatorial direction 
AGNs are classified as Type 2.

Spectropolarimetry is an important tool in testing the unification models 
(e.g.  \citealt{antonucci93}). In the case of the prototypical Seyfert 2 
galaxy NGC~1068, \citet{antonucci85} found the broad H$\beta$ emission 
in the polarized flux spectrum (see also e.g. \citealt{miller90}, \citealt{tran99}). 
They interpreted this broad H$\beta$ by assuming the presence of an electron 
scattering medium in the polar direction that reveals the broad emission line 
region through Thomson scattering.

Another important test of the unification models is provided by
X-ray observations, which reveal that the hardness ratio tends to be 
systematically higher for type 2 AGNs than type 1 AGNs. In particular, 
type 2 AGNs typically 
show severe extinction in the soft X-ray region, indicating the presence
of an absorbing 
component with hydrogen column density of the order of $N_H \sim 10^{23}{\rm\ 
cm^{-2}}$ (e.g. \citealt{ricci14}).  Hard X-ray observations of AGNs also 
show that Compton thick 
material with column densities $>10^{24}{\rm\ cm^{-2}}$ may obscure the AGN 
engines. Compton scattering from high column components is proposed to be an 
important contributor to the cosmic X-ray background that is characterized 
by a broad bump at $\sim 30{\rm\ keV}$  (e.g. see \citealt{comastri15} and \citealt{magdziarz95}).
However, the existence and detailed physical conditions of the molecular 
torus component are only poorly constrained.

As pointed out by \citet{chang15}, a significant amount of neutral hydrogen 
that may exist in the hypothetical molecular torus can act as a Rayleigh
and Raman scattering 
medium. The cross section for Rayleigh scattering increases sharply 
near Ly$\alpha$ due to resonance. 
This implies that far UV radiation near Ly$\alpha$ can be strongly polarized
due to Rayleigh scattering depending on the covering factor 
and the \ion{H}{I} column density
of the molecular torus. This leads to an interesting possibility of polarized 
Ly$\alpha$ to put meaningful
constraints on the unification models of AGNs.

Recent detections of polarized Ly$\alpha$ emission from a
giant Ly$\alpha$ nebula opens a new window to probe the physical
environment of the intergalactic medium illuminated by starbursting
galaxies and active galactic nuclei in the early universe. Ly$\alpha$
nebulae, or ``blobs,'' are rare, extended sources at $z$ = 2--6 with
typical Ly$\alpha$ sizes of 10\arcsec\ ($\sim$100\,kpc) and line
luminosities of $L_{\rm{Ly\alpha}}$ $\sim$ $10^{44}$ erg\,s$^{-1}$
\cite[e.g.,][]{Steidel00, Francis01, Matsuda04, Dey05, Nilsson06, Yang09,
Yang10, Matsuda11, Prescott12}. Recently, \cite{Hayes11} discovered a
polarization pattern of concentric rings within a Ly$\alpha$ nebula at
$z$ = 3.1, which supports a central powering mechanism within the nebula.
While the physical scale of Ly$\alpha$ nebulae ($\sim$100\,kpc) is much
larger than that of AGN central engines ($\sim$ 1 pc), the basic pictures
are the same: Ly$\alpha$ or continuum photons from the central source are
resonantly or Rayleigh scattered by the surrounding neutral \ion{H}{I}
medium.  However, theoretical modeling of the Ly$\alpha$ polarization
is still in its infancy (\citealt{dijkstra08}), and therefore a detailed
modeling of Ly$\alpha$ polarization in diverse geometry is required.

% The high column density component of \ion{H}{I} 
%has also been discussed in the context of spectropolarimetry of H$\alpha$ as a 
%theoretical possibility of broad polarized flux seen in several narrow 
%line AGN (Lee \& Yun 1998). 

We compute, in this paper, the polarization of the far UV radiation 
around Ly$\alpha$  that is Rayleigh scattered in a thick neutral hydrogen
region. The paper is composed as follows.  In section 2, we discuss 
the atomic physics of Rayleigh scattering around Ly$\alpha$.
In the following section, the scattering geometry and our Monte Carlo code 
are described. The main 
result is presented in section 3. In section 4 we discuss the observational 
implications of polarized Ly$\alpha$ in an AGN.

\section{Atomic Physics of Rayleigh Scattering}

\begin{figure}
\includegraphics[scale=0.65]{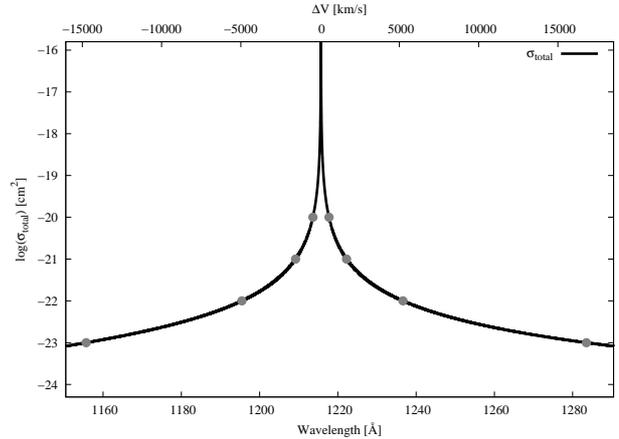}
\caption{Cross section for Rayleigh scattering by atomic hydrogen. The
vertical scale is logarithmic, whereas the horizontal axis is
wavelength in units of \AA. The dots mark the wavelengths
having the cross section values of  $10^{-20},10^{-21},10^{-22},10^{-23}{\rm\
cm^2}$. The cross section exhibits redward asymmetry
as pointed out by \citet{lee13}. }
\label{cross_sect}
\end{figure}

There have been many basic research works on the Rayleigh scattering
processes (e.g. \citealt{isliker89}, \citealt{gavrila67}, 
\citealt{sadeghpour92}, \citealt{lee13}).  The time-dependent second order 
perturbation theory is used to describe
the Rayleigh scattering process. The scattering cross section is  given by
the Kramers-Heisenberg formula, which can be written as
\begin{equation}
{d\sigma_{Ray}\over d\Omega} = r_0^2 \left|(\epsilon^\alpha\cdot\epsilon^
{\alpha'})\left(\sum_n M^b(n)+\int_0^\infty dn' M^c(n')\right)\right|^2.
\end{equation}
Here, $\epsilon^\alpha$ and $\epsilon^{\alpha'}$ are the polarization 
vectors associated with the incident and outgoing photons, respectively,
and $r_0$ is the classical electron radius.
The angular integral over the two polarization states for an incident 
unpolarized radiation results in the factor of $8\pi/3$, which is in turn 
multiplied by $r_0^2$ to yield the Thomson scattering cross section 
$\sigma_T=0.665\times10^{-24}{\rm\ cm^2}$.  In the formula, the matrix elements 
between the ground $1s$ state and the bound $np$ states are given by
\begin{equation}
M^b(n) = {m\omega^2\over 3\hbar}|<1s\parallel r\parallel np>|^2
{2\omega_{n1}\over \omega_{n1}^2-\omega^2} ,
\end{equation}
and those between $1s$ and continuum $n'p$ states are given by
\begin{equation}
M^c(n') = {m\omega^2\over 3\hbar}|<1s\parallel r\parallel n'p>|^2
{2\omega_{n'1}\over \omega_{n'1}^2-\omega^2}.
\end{equation}
The explicit expressions of the reduced matrix elements can be found in many text 
books on quantum mechanics
(e.g. \citealt{berestetskii71}, \citealt{saslow69}). 
%\begin{eqnarray}
%|<1s\parallel r\parallel np>|^2 &=&  {2^8 n^7 (n-1)^{2n-5}
% \over (n+1)^{2n+5}}
%\nonumber \\
|%<1s\parallel r\parallel n'p>|^2 &=& {2^8 n'^7 \exp[-4n'\tan^{-1}(1/n')]
%\over (n'^2+1)^5[1-\exp(-2\pi n')]}
%\end{eqnarray}
The Rayleigh scattering cross section is shown in Fig.~\ref{cross_sect},
where the horizontal axis is wavelength in units of \AA\ and the
vertical scale is the logarithm of the cross section in units of cm$^2$. 
A useful approximation to the Rayleigh scattering cross section can be also
found in \citet{lykins15}.

In particular, near Ly$\alpha$,  the cross section is dominantly
contributed by the first term in the summation leading to an approximation
\begin{equation}
\sigma_{Ray}(\lambda) \simeq
\sigma_T \left[ {f_{\alpha}\over (\lambda/\lambda_\alpha)-1} \right]^2,
\end{equation}
where $f_\alpha=0.4162$ is the oscillator strength for the Ly$\alpha$ transition and
$\lambda_\alpha=1215.67 \,$\AA\ is the Ly$\alpha$ wavelength. 

Given a neutral column density $N_{HI}$, we may consider the scattering band width 
$\Delta \lambda_R(N_{HI}=\lambda-\lambda_\alpha$ 
by the wavelength width around Ly$\alpha$, for which 
the Rayleigh scattering optical depth exceeds unity. 
Corresponding to $\Delta\lambda_R(N_{HI}$, we have the Doppler velocity width
$\Delta V=c (\lambda- \lambda_\alpha)/\lambda_\alpha $. The requirement
$\sigma_{Ray}(\lambda)N_{HI}=1$ leads to
\begin{equation}
\left({\Delta V\over c}\right)^2 \simeq \sigma_T N_{HI} c^2 f_{12}^2,
\end{equation}
or numerically
\begin{equation}
{\Delta V\over 10^4\ {\rm km\ s^{-1}}}\simeq {\left[N_{HI}\over
3\times 10^{22}\ {\rm cm^{-2}}\right]^{1/2}}.
\end{equation}

 Because the dynamical velocity scale of the
broad line region of a typical AGN is about $10^4\ {\rm km\ s^{-1}}$,
most of the broad Ly$\alpha$ line photons can be Rayleigh scattered
under the assumption that $N_{HI} \ge 3 \times 10^{22}\ {\rm cm^{-2}}$.  
This velocity scale is much larger
than the thermal speed of typically photoionized emission nebulae, no
consideration is made of resonance scattering of Ly$\alpha$ in this work
(e.g. \citealt{dijkstra08}). 

The dots in Fig.~\ref{cross_sect} correspond to those wavelengths with 
Rayleigh scattering optical thickness $\tau_s$ from $0.1$ to $100$
for \ion{H}{I} column density of $N_{HI}=10^{21}{\rm\ cm^{-2}}$.
As is noted in these values, the asymmetry in the Rayleigh cross
section is quite significant (see also \citealt{totani16} and
\citealt{bach14}). Therefore we expect that the effect of asymmetry in scattering cross section
will be apparent in the flux profile and polarization
of Rayleigh scattered radiation. 

In this work, we consider far UV radiation in the fixed wavelength range
$\lambda_{\alpha} - 70$\ \AA\ $<\lambda<$ $\lambda_{\alpha} + 70$\ \AA,
and the cross section at the lower and upper wavelength limits is 
$\sim 10^{-25}{\rm\ cm^2}$.

\section{Monte Carlo Simulations of Rayleigh Scattering}

\subsection{Scattering Geometry}

\begin{figure*}
\includegraphics[scale=0.45]{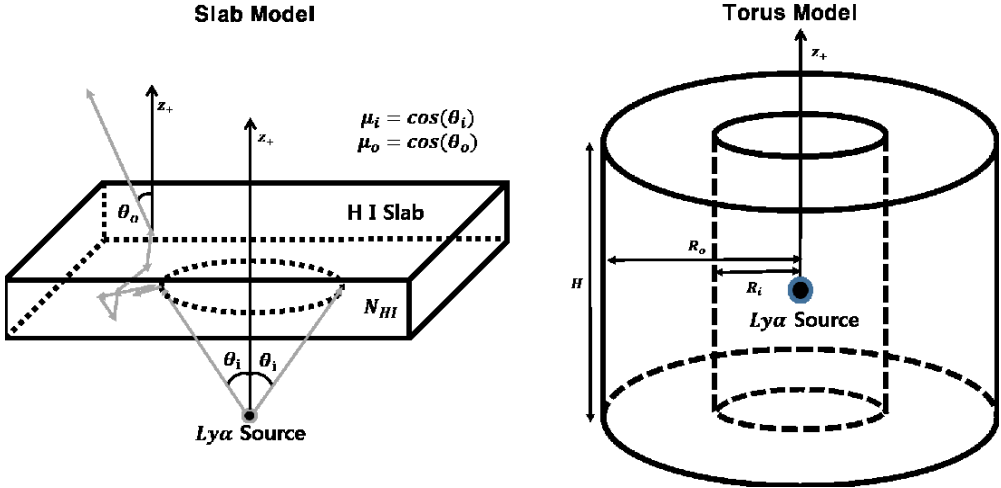}
\caption{Schematic illustration of the scattering geometry adopted in this 
work. The Ly$\alpha$ emission source is located at the center of 
the coordinate system. In the case of the slab geometry, hydrogen atoms 
are uniformly distributed between $z=z_0>0$ and $z=z_0+H$, where $H$ is 
the thickness of the slab. The column density $N_{HI}$ is measured 
along the $z$-direction
in the slab geometry. In the torus geometry, hydrogen atoms are
uniformly distributed inside a cylinder shell characterized by 
the inner and outer radii $R_i$ and $R_o=2R_i$ and the height $H=AR_i$.
% The inset shows the spectrum of the incident radiation, which can be
% decomposed into the flat continuum and the broad Ly$\alpha$ emission
% with a triangular profile.
}
\label{geometry}
\end{figure*}

%\begin{figure}
%\includegraphics[scale=0.69]{input_bw.ps}
%\caption{Input Spectra
%}
%\label{input}
%\end{figure}

In this section, we illustrate the scattering geometry adopted in this work. 
We consider a neutral hydrogen region in the shape of a slab with a finite 
thickness and infinite lateral extension. The slab geometry is a convenient 
model to study the fundamental properties of the polarization of Rayleigh 
scattered radiation and verify the Monte Carlo code as well.
In this case, the \ion{H}{I} column density $N_{HI}$ along the normal 
direction completely specifies the scattering geometry.

The other scattering geometry considered in this work is a finite cylindrical 
shell which is characterized by the height $H$ and the
inner and outer radii $R_i$ and $R_o$, respectively. In this work, 
we set the parameters $R_o=2R_i$ so that the inner radius is the same 
as the lateral width of the shell. There are two controlling parameters 
for this geometry. One is the \ion{H}{I} column density $N_{HI}=n_{HI}(R_o-R_i)$ in the lateral direction
and the other is the shape parameter given by the ratio $A=H/R_i$.

 A schematic illustration of our scattering geometry
is shown in Fig.~\ref{geometry}. We set the coordinate system so 
that $z$-axis coincides with the symmetry axis
and that the Ly$\alpha$ emission source is located at the origin.
 In our Monte Carlo simulations
we collect emergent photons according to the $z$ component $\mu_o={\rm cos}\theta_o$
of the unit wavevector ${\bf\hat k}$ with the bin size of $\Delta\mu_o=0.1$. 
In the slab case, the Ly$\alpha$
emission source is located on the negative $z$-axis, so that
we will refer emergent photons with negative and positive $\mu_o$ to Rayleigh 
reflected and transmitted photons, respectively.

In this work, we consider two types of incident radiation, one consisting 
of pure flat continuum and the other with an additional contribution 
of broad Ly$\alpha$ emission with a triangular profile. 
The triangular profile is chosen for simplicity of interpretation.
The equivalent width of the Ly$\alpha$ emission is set to
be 90 {\AA} and the FWHM (full width at half maximum) of that is 20 {\AA}, $\simeq 5000{\rm\ km/s}$ in the scale of speed,
which appears to be typical in many AGNs \citep{vanden01}.

The density matrix formalism is adopted in our Monte Carlo code
to describe the polarization information. 
A $2\times$2 density matrix $\rho$ is  defined by
\begin{equation}
\rho=\left(\begin{matrix} (I+Q)/2 & (U+iV)/2 \\
(U-iV)/2 & (I-Q)/2 \\ \end{matrix} \right)
\end{equation}
in terms of the Stokes parameters $I, Q, U$ and $V$
(see e.g. \citealt{ahn15}).

In this simulation the two polarization basis vectors associated with the unit
wavevector ${\bf\hat k}=(\sin\theta\cos\phi,\sin\theta\sin\phi, \cos\theta)$
are chosen as
\begin{eqnarray}
{\epsilon}_1 &=& (-\sin\phi,\cos\phi,0) 
\nonumber \\
{\epsilon}_2 &=& (\cos\theta\cos\phi,\cos\theta\sin\phi,-\sin\theta)
\end{eqnarray}
so that the ${\epsilon}_1$ and ${\epsilon}_2$ represent the polarization in the direction perpendicular and parallel to the symmetry axis, respectively.

For each photon generated in the simulation, a density
matrix with a unit trace is assigned and followed until escape.
Being characterized by the density matrix elements $\rho_{11}=\rho_{22}=0.5,
\rho_{12}=\rho_{21}=0$, the initial photon from the source is assumed to be 
completely unpolarized. 

According to \citet{schmid95} the polarization of Rayleigh scattered 
radiation is characterized by the same phase function as that of Thomson 
scattering. The unnormalized density matrix $\rho_p'$ associated 
with the scattered photon with a new unit wavevector
${\bf\hat k}=(\sin\theta'\cos\phi',\sin\theta'\sin\phi', \cos\theta')$ is 
computed using the equation
\begin{equation}
(\rho_p')_{ij}=\sum_{kl=1,2}(\epsilon'_i\cdot\epsilon_k)\rho_{kl}(\epsilon_l\cdot\epsilon_j').
\end{equation}
The components are explicitly written as
\begin{eqnarray}
(\rho_p')_{11} &=& (\cos^2\Delta\phi)\rho_{11}
\nonumber\\
                &-& (\cos\theta \sin2\Delta\phi)\rho_{12}
\nonumber\\
                &+& (\sin^2\Delta\phi \cos^2\theta)\rho_{22}
\nonumber\\
(\rho_p')_{12} &=& ({1 \over 2} \cos\theta'\sin2\Delta\phi)\rho_{11}
\nonumber\\
                &+& (\cos\theta \cos\theta' \cos2\Delta\phi + \sin\theta \sin\theta' \cos\Delta\phi)\rho_{12}
\nonumber\\
                &-& \cos\theta(\sin\theta \sin\theta' \sin\Delta\phi + {1 \over 2} \cos\theta \cos\theta' \sin2\Delta\phi)\rho_{22}
\nonumber\\
(\rho_p')_{22} &=& (\cos^2\theta' \sin^2\Delta\phi)\rho_{11}
\nonumber\\
                &+& \cos\theta'(2\sin\theta \sin\theta' \sin\Delta\phi + \cos\theta \cos\theta' \sin2\Delta\phi)\rho_{12}
\nonumber\\
                &+& (\cos\theta \cos\theta' \cos\Delta\phi + \sin\theta \sin\theta')^2 \rho_{22} 
\end{eqnarray}

The trace of the unnormalized density matrix represents the angular 
distribution of the scattered radiation, from which 
we select $\theta'$ and $\phi'$ in a probabilistic way (Lee et al. 1994).
Once the selection is made, the density matrix is normalized to have a unit 
trace. In the density matrix formalism adopted in this work, the off diagonal 
element $\rho_{12}$ remains real because Rayleigh scattering induces 
no circular polarization for an incident radiation with no circular 
polarization. In other words, the density matrix in this work
is a 2$\times$2 real symmetric matrix. Furthermore, due to the axial
symmetry of the scattering geometry considered in this work, linear 
polarization can develop
only in the direction parallel or perpendicular to the symmetry axis when spatially averaged, leading to
the vanishing average of $\rho_{12}$.
The resultant linear polarization $Q$ is represented by the difference of the two diagonal elements 
of $\rho$ with a unit trace,
\begin{equation}
Q=\rho_{11}-\rho_{22}.
\end{equation}

Here, a positive $Q$ implies that the polarization develops in the direction perpendicular
to the symmetry axis, whereas a negative $Q$ shows polarization in the parallel direction.
The scattering geometries considered in this work possess axis-symmetry and our
choice of the Stokes parameter $Q$ to represent polarization along the symmetry axis makes
the Stokes parameter $U$ vanish.

\subsection{Rayleigh Scattering in a Slab Region}

\begin{figure}
\includegraphics[scale=0.75]{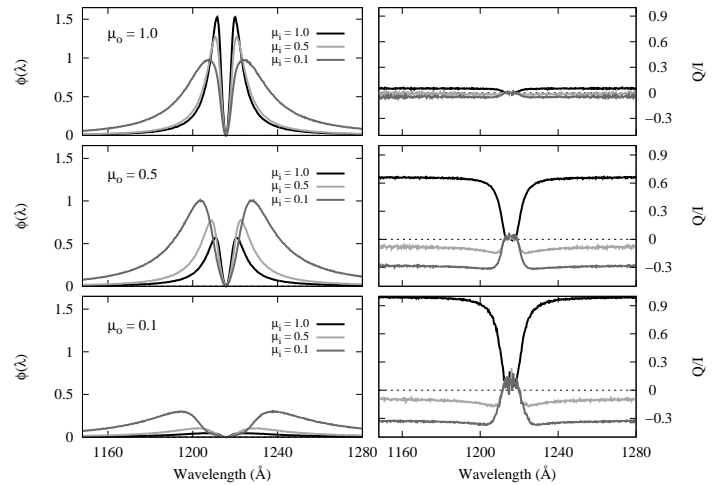}
\caption{Degree of linear polarization (right panels) and flux (left panels) 
of far UV flat continuum around Ly$\alpha$ 
that are Rayleigh transmitted from a slab with the
\ion{H}{I} column density $N_{HI}=10^{21}{\rm\ cm^{-2}}$ in the normal direction.
The upper, middle and bottom panels are for those photons emergent with $\mu_o$
in the interval $(0.9,1)$, $(0.4,0.5)$ and $(0,0.1)$, respectively. In the right panels,
a positive $Q/I$ implies polarization in the direction perpendicular to the slab normal.
}
\label{spec_tran_slab_flat}
\end{figure}

\begin{figure}
\includegraphics[scale=0.75]{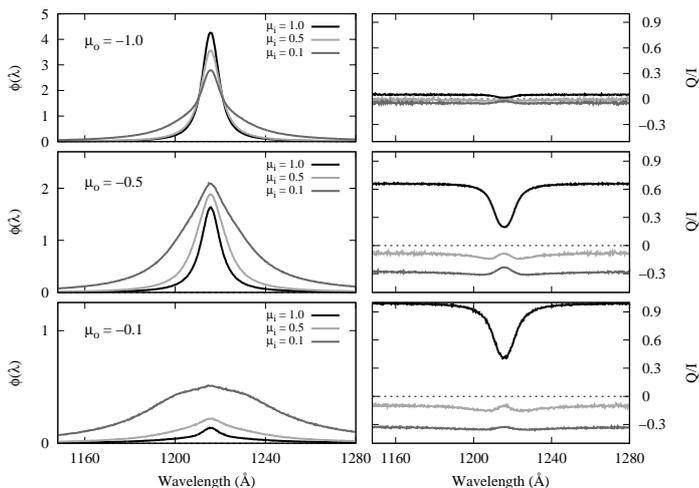}
\caption{Polarization and flux of far UV flat continuum around Ly$\alpha$ 
Rayleigh reflected from the same slab considered in Fig.~\ref{spec_tran_slab_flat}.
The upper, middle and bottom panels are for those photons emergent with $\mu_o$
in the interval $(-1,-0.9)$, $(-0.5,-0.4)$ and $(-0.1,0)$, respectively.
}
\label{spec_ref_slab_flat}
\end{figure}

In this section, we present our Monte Carlo results for Rayleigh
scattering in a slab.
Note that in sections 3.2.1 and 3.2.2, 
we only collect photons that are scattered at least once and 
neglect optically thin photons that pass through the slab without any interaction
because our goal is to investigate the polarization properties of the Rayleigh scattered
photons depending on the $\mu_o$, $\mu_i$, and $N_{HI}$.

In section 3.2.3, we integrate over all $\mu_i$ to investigate the flux and polarization
of radiation originated from an isotropic source.  We also generate
mock spectra using a flat continuum source and a emission source with a triangular
profile.

\subsubsection{Spectra of Reflected and Transmitted Radiation}

In Fig.~\ref{spec_tran_slab_flat}, we show the flux (left panels) 
and polarization degree (right panels) for Rayleigh transmitted radiation 
from a slab with the \ion{H}{I} column density of $N_{HI}=10^{22}{\rm\ cm^{-2}}$ 
in the normal direction.  The far UV incident radiation is to be taken 
as a flat continuum, characterized by the same number of photons emitted 
with specified $\mu_i$ per unit wavelength interval.
Here $\mu_i$ represent the $z$-component of the unit vector in the direction
of the incident photons.
The upper panels are flux and degree of polarization of photons emergent 
with $\mu_o$ between 0.9 and 1.0. The middle and bottom panels show 
the same information for $\mu_o$ in the intervals (0.4,0.5) 
and (0,0.1), respectively.

The left panels show the flux profiles for various $\mu_i$ and $\mu_o$. 
All the transmitted profiles are characterized by the central dip 
of which the width increases as $\mu_o$ decreases from 1 to 0. This implies 
that the central dip reflects the \ion{H}{I} column density of the slab 
in the direction of incidence. On the other hand, the much 
extended wing features delineate the scattering cross section that is 
approximated by a Lorentzian.
Being optically thin, most far UV radiation in this wavelength regime will pass through the slab 
without any interaction with atomic hydrogen.

When $\mu_o\sim 1$, the degree of polarization is nearly zero due to the axial 
symmetry. This is illustrated in the right top panel, where irrespective 
of $\mu_i$, $Q/I$ is negligibly small.  This case may serve as a check 
of our code.

An opposite case can be seen in the right bottom panel. For $\mu_i\sim 1$ 
and $\mu_o\sim 0.1$ corresponding to the normal incidence and emergence 
in the grazing direction, the degree of polarization in the wing region is 
nearly 1 in the direction perpendicular to the slab normal ($z$ axis in Fig.\ref{geometry}). However, 
near line center the polarization becomes quite weak due to the
sharp increase of scattering optical depth.

It is interesting to observe that $\mu_i\simeq 0.1$ and $\mu_o\sim 0.1$
 the polarization develops in the direction parallel to the symmetry axis 
with a typical degree of polarization about 30 \%
in the wing part. 
This is because most photons are singly scattered at the large distance 
from the origin, thus strongly polarized in a direction parallel to the symmetry axis 
given that the wavevectors of the incident and transmitted photons are 
located within $z\sim0$ plane.

In this case also the polarization becomes rapidly weak 
as the scattering optical depth increases toward the line center. 
For $\mu_i\simeq 0.5$ and $\mu_o\simeq 0.1$, the polarization is very weak 
because it corresponds to an intermediate case of the two cases
of the normal and grazing incidences.
The middle panels show the intermediate behavior between that 
of the case $\mu_o=1$ shown in the top panels and that for $\mu_o=0.1$.

In Fig.~\ref{spec_ref_slab_flat}, we show our results for Rayleigh
reflected radiation near Ly$\alpha$ from the same slab considered
in Fig.~\ref{spec_tran_slab_flat}. In the left panels showing the flux
as a function of wavelength, we immediately note that all the
profiles are characterized by a single broad peak of which the width
is dependent on the incident and emergent directions. The photons
constituting the central peak are optically so thick that they rarely
penetrate to contribute to the transmitted flux considered in 
Fig.~\ref{spec_tran_slab_flat}.

In the right panels, we show the degree of linear polarization
of the Rayleigh reflected radiation. In the top panel, the
radiation is negligibly polarized because the observer's line
of sight nearly coincide with the symmetry axis.
An interesting point can be noted in the middle and bottom
panels for $\mu_i=1.$ In these cases the center part is the most
weakly polarized but still exhibits fairly high degree of linear 
polarization in excess in 10 per cent. This is highly contrasted
with the negligibly polarized
cases of the transmitted flux shown in the middle and bottom 
panels of Fig.~\ref{spec_tran_slab_flat}.

This contrasting behavior can be understood by considering the
fact that the reflected flux is quite significantly contributed by
the singly Rayleigh scattered near the point of entry.
These singly scattered photons are highly polarized in the direction
perpendicular to the symmetry axis. However, near the
Ly$\alpha$ center in the transmitted flux
we expect no such singly scattered photons due to the huge
scattering optical depths. This leads to negligibly polarized dip
in the transmitted flux.

Another interesting point to be noted is the fact that we have
a fairly constant degree of linear polarization for the cases
of oblique and grazing incidence where $\mu_i\le 0.5$. This
phenomenon is also attributed to the fact that most contribution
is made by photons scattered only a few times irrespective of
wavelength.

\subsubsection{Effects of \ion{H}{I} Column Density}

%\begin{figure}
%\includegraphics[scale=0.69]{tflux_slab_flat_bw.ps}
%\caption{Total flux from a slab and continuum.
%}
%\label{tflux_slab_flat}
%\end{figure}

\begin{figure}
\includegraphics[scale=0.69]{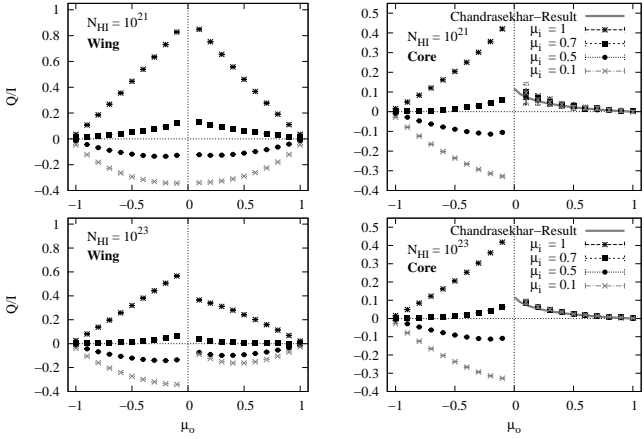}
\caption{Polarization behaviors for wing photons $\tau_s<10$ 
Rayleigh scattered from a slab (left panels) and for 
core photons $\tau_s>10$ (right panels). In order to find the
dependence on \ion{H}{I} density, we consider two cases
of $N_{HI}=10^{21}{\rm\ cm^{-2}}$ (top panels) and
$N_{HI}=10^{23}{\rm\ cm^{-2}}$. The horizontal axis shows
$\mu_0$, where $\mu_o<0$ is for Rayleigh reflected radiation
and $\mu_o>0$ for Rayleigh transmitted radiation.
}
\label{tpol_slab_flat}
\end{figure}

In Fig.~\ref{tpol_slab_flat}, we show the polarization behavior 
as a function of emergent direction for two values of
$N_{HI}=10^{21}{\rm\ cm^{-2}}$ (top panels) 
and $10^{23}{\rm\ cm^{-2}}$(bottom panels).  The horizontal axis shows
$\mu_0$, where $\mu_o<0$ is for Rayleigh reflected radiation
and $\mu_o>0$ for Rayleigh transmitted radiation. The different
symbols represent various values of $\mu_i$.

We also divide the
incident radiation into two parts depending on the scattering optical thickness
so that the left panels show the emergent photons for $\tau_s <10$ (wing) 
and the right panels for $\tau_s>10$ (core).
The wavelengths corresponding to $\tau_s=10$ 
for $N_{HI}=10^{21}{\rm\ cm^{-2}}$
are $\lambda_1 = 1213.61$~{\rm \AA} and $\lambda_2 = 1217.74$~{\rm \AA}.
For $N_{HI}=10^{23}{\rm\ cm^{-2}}$ the corresponding
wavelengths are $\lambda_3=1195.45$~{\rm \AA} and $\lambda_4=1236.62$~{\rm \AA}.

Because the incident radiation is prepared in the wavelength 
range $\lambda_{\alpha}-70${\rm\ \AA}$< \lambda 
< \lambda_{\alpha}+70${\rm\ \AA} throughout all the Monte Carlo simulations
presented in this work, the larger
wavelength range for $\tau_s<10$ for $N_{HI}=10^{21}{\rm\ cm^{-2}}$ than for $N_{HI}=10^{23}{\rm\ cm^{-2}}$ implies
%that more optically thin photons are 
%incident on the slab in the case of top left panel
%than in the case of bottom left. 
that the slab in the case of top left panel is optically thin to
more incident photons than in the case of bottom left.
The Rayleigh scattering
phase function is forward and backward symmetric so that
an optically thin slab shows the same polarization behavior
for transmitted and reflected components. This is illustrated
in the symmetric polarization with respect to $\mu_o=0$
in the top left panel. In the bottom left panel for
$N_{HI}=10^{23}{\rm\ cm^{-2}}$ symmetry is significantly broken due
to more contribution from photons with optical depths exceeding unity.
These photons with optical depths greater than unity
will be scattered many times to become quite weakly polarized.
This tendency becomes more severe for photons emergent near
the grazing direction ($\mu_o = 0$). Therefore, optically thin
Rayleigh reflected radiation is maximally polarized in the grazing
direction, whereas it is no longer the case for
Rayleigh transmitted radiation from grazing incidence ($\mu_i = 0$).

This observation is confirmed in the behaviors shown in the right
panels, where all the photons are characterized by $\tau_s>10$.
Rayleigh reflected components are significantly contributed by
photons with a small number of scatterings leading to strong
polarization. Maximal asymmetry is obtained for photons emergent
in the grazing direction, which leads to maximum degree of polarization. 
This also explains again the stronger polarization
shown in the spectra of Rayleigh reflected radiation than
in those of Rayleigh transmitted radiation illustrated
in Figs.~\ref{spec_tran_slab_flat}
and \ref{spec_ref_slab_flat}.

A very interesting behavior of Rayleigh transmitted radiation 
is recognized in the right panels, where the scattering optical
depths are very large.
The polarization develops in the direction perpendicular
to the slab normal independent of $\mu_i$. 
Furthermore, the degree of polarization increases
as $\mu_o$ approaches 0 from 1. The limit value is 11.7 \%,
which was obtained by \citet{chandrasekhar60} in the polarized
transfer of Thomson scattered radiation in an infinitely thick slab
(see also \citealt{angel69}).
A similar result was obtained by \cite{ahn15} in their study
of the polarized transfer of resonantly scattered Ly$\alpha$.

\subsubsection{Polarized Ly$\alpha$ from an Isotropic Source}

\begin{figure}
\includegraphics[scale=0.69]{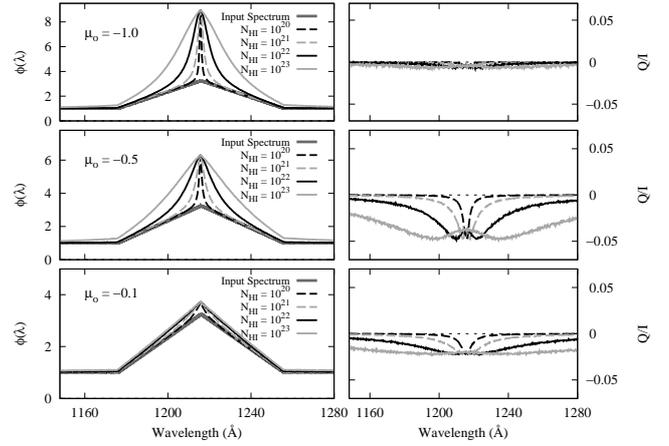}
\caption{Rayleigh reflected degree of polarization(right panel) and flux(left panel) of the isotropic source composed the flat continuum and broad emission. The various lines are shown by H I column densities $N_{HI}$ of slab model.
}
\label{spec_ref_slab_mock}
\end{figure}

\begin{figure}
\includegraphics[scale=0.69]{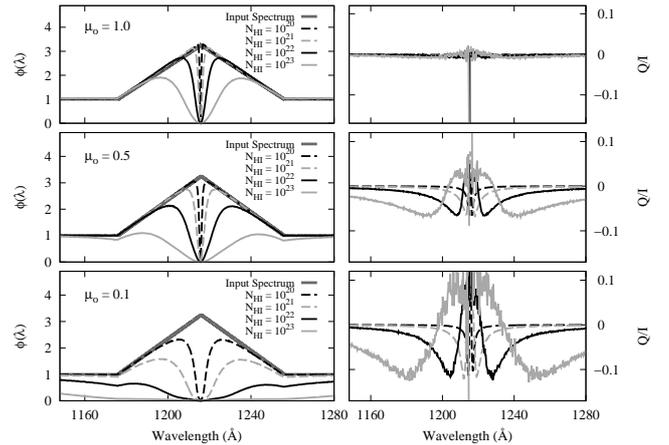}
\caption{Transmitted component of same model as Fig.~\ref{spec_ref_slab_mock}.
}
\label{spec_tran_slab_mock}
\end{figure}

%\begin{figure}
%\includegraphics[scale=0.69]{tpol_tflux_slab_mock_bw.ps}
%\caption{Total flux and polarization from a slab and emission + continuum.
%}
%\label{tpol_tflux_slab_mock}
%\end{figure}

In previous sections, we considered photons only scattered at least once by hydrogen atoms, 
in which case non-scattered radiation would reduce the observed polarization significantly.
In contrast in this section we consider an isotropic source and collect all the photons
including those without being Rayleigh scattered.
In Figs.~\ref{spec_ref_slab_mock} and \ref{spec_tran_slab_mock},
%and \ref{tpol_tflux_slab_mock}, 
%we combine the results shown
%in Figs.~\ref{spec_ref_slab_tra} and \ref{spec_tran_slab_tra}
%and \ref{tpol_slab_tra} 
we show the polarization 
of Rayleigh scattered radiation in a slab illuminated by an isotropic
source for 4 different values of $N_{HI}$. The input spectrum
is the sum of a flat continuum and a broad Ly$\alpha$ emission line
flux with a triangular profile described in section~3.1.
The top panel of Fig.~\ref{spec_ref_slab_mock} shows the Rayleigh reflected
component emergent in the normal direction, which is negligibly polarized
due to symmetry.

Because the covering factor of the scattering region with respect to
the emission source is 50 \%, the emergent flux is significantly
distorted in the wavelength region with $\tau_s>10$ leading to
the formation of the sharp central feature for low values of $N_{HI}$.
As shown in the bottom left panel, the Rayleigh reflected in the
grazing direction shows almost the same profile as the incident radiation.
This is due to large $\tau_s$ in this direction, giving rise to 
full reflection of the incident radiation.

Rayleigh reflected radiation in a slab is polarized in
the direction parallel to the symmetry axis, which is illustrated by the
negative degree of linear polarization in the right panels
of Fig.~\ref{spec_ref_slab_mock}.  Strong polarization
is obtained in the case of $\mu_o=-0.5$ and weak polarization is
seen for the Rayleigh reflected radiation emergent in the grazing
direction. In the far wing region where the scattering optical
thickness is very small, we obtain very weak polarization.
It is notable that there exist local maxima in the degree of
linear polarization (minima in $Q/I$ value) around the shoulder region for $\mu_o=-0.5$.
Here, the term 'shoulder region' is meant to indicate the intermediate region
between the core and wing parts.
This is explained by the fact that the line center is contributed
more significantly by photons scattered many times and hence polarized 
very weakly than the shoulder region is.

The left panels of Fig.~\ref{spec_tran_slab_mock}
show the spectra of Rayleigh transmitted radiation from an isotropic
source with a triangular broad Ly$\alpha$ emission. As is previously
pointed out, the spectra are characterized by the central dip that 
depends on the emergent direction and $N_{HI}$.
In the middle right panel and the bottom right panel, the
degree of polarization exhibits a complicated behavior.
The degree of polarization is positive near line center
and negative from the shoulder and wing regions.

This complicated behavior can be understood as follows.
For a very optically thick photon the radiative transfer through a thick slab 
must involve a diffusive propagation along the vertical $z-$ direction until it reaches
a slab boundary before escape.  The diffusive propagation along $z-$direction induces development of polarization 
in the direction perpendicular to $z-$ axis. For photons emergent in the grazing
direction, the degree of polarization can reach 11.7 per cent,
which was shown by \citet{chandrasekhar60} and illustrated
in the bottom right panel of Fig.~\ref{spec_tran_slab_mock}.
However, in the far wing regions where photons are optically thin,
scattering plane nearly coincides with the slab plane leading to
polarization in the direction parallel to $z-$ axis.
An analogous phenomenon can be found in the case of Thomson 
scattering investigated by \citet{phillips86}.
We point out that the polarization flip around Ly$\alpha$ can be an important  signature
of Rayleigh transmitted Ly$\alpha$ in a thick neutral slab and that the wavelength at which
the flip occurs indicates the characteristic neutral column density of the slab.

%\subsubsection{Polarization of Rayleigh Thin and Thick Radiation}

\begin{figure}
\includegraphics[scale=0.69]{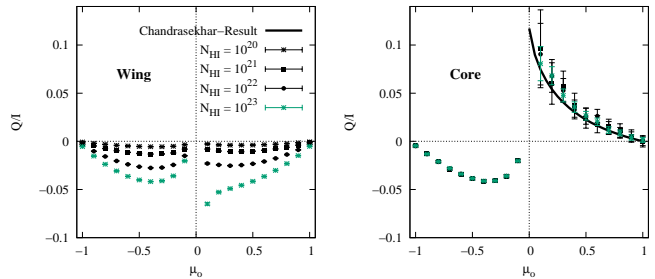}
\caption{Degree of polarization for wing photons(left panel) and core photons(right panel).
$\mu_o<0$ is for reflected component and $\mu_o>0$ is for transmitted component. 
}
\label{tpol_slab_mock}
\end{figure}

The left and right panels of Fig.~\ref{tpol_slab_mock} show the polarization 
behaviors of radiation with low and high Rayleigh scattering optical depths, 
respectively. In this figure, the simulation parameters are the same as 
in Figs.~\ref{spec_ref_slab_mock} and \ref{spec_tran_slab_mock}.

In the left panel, the wing photons both Rayleigh reflected
and transmitted are polarized in the direction parallel to the
symmetry axis. For small values of $N_{HI}\le 10^{21}{\rm\ cm^{-2}}$ 
the polarization behaviors are similar for Rayleigh reflected and transmitted 
components.  However, for a very thick $N_{HI}=10^{23}{\rm\ cm^{-2}}$ 
the degree of polarization peaks at $\mu_o\simeq 0$ for Rayleigh transmitted 
radiation, because they are significantly contributed by singly scattered 
photons.

In the right panel we notice that the Rayleigh reflected radiation is
always polarized in the direction parallel to the symmetry axis irrespective of
the Rayleigh scattering optical depth. However, Rayleigh transmitted radiation
can be polarized in the direction
perpendicular to the symmetry axis only when the Rayleigh scattering
optical depth is very high.

\begin{figure}
\includegraphics[scale=0.45]{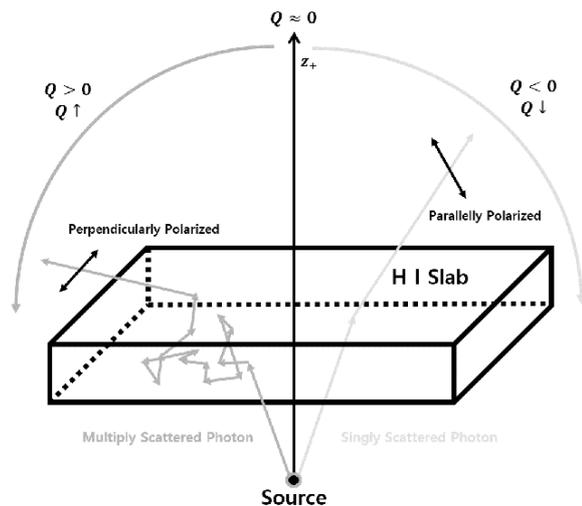}
\caption{Description of degree of polarization for optically thin and thick cases in the slab model.
}
\label{slab_polarization}
\end{figure}

In Fig.~\ref{slab_polarization}, we schematically illustrate polarization behaviors of Rayleigh scattered photons
in optically thin and thick cases.
Multiply scattered photons tend to be  polarized in the direction perpendicular to the symmetry axis as a result
of diffusive propagation along $z-$axis, whereas
singly scattered photons are polarized in the direction parallel to $z-$axis. 
%Through this figure we can understand polarization of Rayleigh scattered Ly$\alpha$.

%\begin{figure*}
%\includegraphics[scale=0.45]{Geometry.ps}
%\caption{Schematic illustration of the scattering geometry adopted in this 
%work. The Ly$\alpha$ emission source is located at the center of 
%the coordinate system. In the case of the slab geometry, hydrogen atoms 
%are uniformly distributed between $z=z_0>0$ and $z=z_0+H$, where $H$ is 
%the thickness of the slab. The column density $N_{HI}$ is measured 
%along the $z$-direction
%in the slab geometry. In the torus geometry, hydrogen atoms are
%uniformly distributed inside a cylinder shell characterized by 
%the inner and outer radii $R_i$ and $R_o=2R_i$ and the height $H=AR_i$.
% The inset shows the spectrum of the incident radiation, which can be
% decomposed into the flat continuum and the broad Ly$\alpha$ emission
% with a triangular profile.
%}
%\label{geometry}
%\end{figure*}

\subsection{Rayleigh Scattering in a Torus Region}

In this section, we consider Rayleigh scattered radiation in a cylindrical
shell region, which approximates the torus geometry that is invoked in the
unification model of AGNs. The torus geometry is symmetric with respect to
the plane $z=0$, and hence the discussion is limited to radiation emergent with 
$\mu_o>0$. In this work, the shape of the cylindrical shell is
described by the parameter 
$A=H/R_i$ defined as the ratio of the height and the inner radius.

\subsubsection{Spectra and Polarization of Flat Continuum}

\begin{figure}
\includegraphics[scale=0.69]{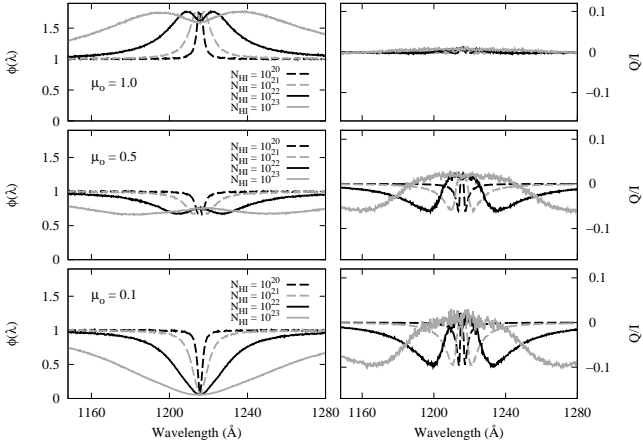}
\caption{Flux and degree of polarization of flat continuum incident on
and Rayleigh scattered in a torus with the shape parameter $A = 2$.
%For $\mu_o\sim 0.5$
}
\label{spec_torus_flat}
\end{figure}

\begin{figure}
\includegraphics[scale=0.69]{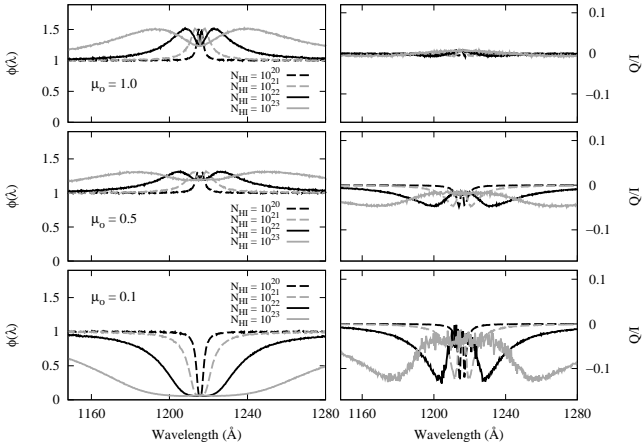}
\caption{Same data as in Fig.\ref{spec_torus_flat} except for $A=1$.
No polarization flip occurs in this case.
}
\label{spec_torus_flat_A1}
\end{figure}

In Fig.~\ref{spec_torus_flat} we show the flux and polarization of Rayleigh
scattered radiation emergent from a torus with $A=1$  for 4 different
values of the \ion{H}{I} column density.  The incident radiation is
pure flat continuum. The top panels show the spectra and degree
of polarization emerging along the symmetry axis (analogous to type 1 AGNs). Because of symmetry
the emergent radiation is nearly unpolarized. The flux profile
exhibits local blue and red maxima. In particular, for 
$N_{HI}=10^{23}{\rm\ cm^{-2}}$ the maxima are seen at $\lambda=1187$~{\rm \AA} and $\lambda=1242$~{\rm \AA}. At these wavelengths
the scattering optical depth is near unity and the whole
scattering region contributes to the flux emergent in the direction
near $z-$axis.

The middle panels show the simulated data collected for photons
emerging with $\mu_o=0.5$.
As is seen in the left panel, the central part is absorbed because
the observer's line of sight is blocked by the cylindrical shell.
At the line center we notice a small mound of Rayleigh scattered flux
so that the flux minima are found at positions shifted redward and blueward
of the line center. 
The formation of the central mound is attributed to a large number
of bouncing near the inner surface of the cylindrical shell by
those photons with huge scattering optical depth $\tau_s$.
Through repeated bouncings they climb up the inner cylindrical shell
until they reach the upper part where direct escape to the line of sight with
$\mu_o=0.5$ is readily made. Therefore,
these photons are polarized in the direction perpendicular to the symmetry
axis. This explains the positive degree of polarization shown by those
photons in the vicinity of the central mound illustrated in the right panel.
In the far wing regions where photons are optically thin, the polarization
develops in the direction parallel to the symmetry axis.

The bottom panel shows the spectra and polarization for grazingly emergent
radiation (analogous to type 2 AGNs). In this case we observe a simple central dip in the left panel because
of the negligible area in the upper part of the cylindrical shell
where direct escape is possible. Polarization near the line center
is very weak and hence significant polarization along the symmetry axis 
can be observed in the wing regions.

In Fig.~\ref{spec_torus_flat_A1}, we plot the same Monte Carlo result 
for a torus with $A=1$. In the middle panel of this figure, we notice
that the polarization develops in the direction parallel to the symmetry axis
in the entire range of wavelength,
and no polarization flip is seen. This is  in high contrast with the behavior
shown in the middle panel of Fig.~\ref{spec_torus_flat}.
The torus geometry with a low value of $A=1$ allows mainly Rayleigh scattering
in the plane nearly perpendicular to the symmetry axis, resulting in polarization
developing in the direction parallel to the symmetry axis. 
The presence of polarization flip in Rayleigh scattered Ly$\alpha$ may indicate
quite significant covering factor of the molecular torus in AGNs.

In the bottom panel we notice the polarization behavior of the grazingly emergent
radiation qualitatively similar to that found in the middle panel. 
For photons with Rayleigh scattering optical depth $\tau_s$ near unity, 
single Rayleigh scattering dominates in the plane nearly coinciding with
the $x-y$ plane, which leads to strong polarization in the parallel direction 
with the degree of polarization in excess of 10 percent. The weak polarization
near the line center shows the effect of multiple scatterings that tend to
randomize the electric field associated with the scattered radiation. 

\begin{figure}
\includegraphics[scale=0.35]{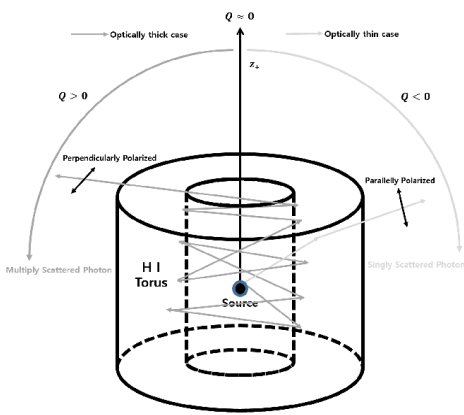}
\caption{Illustration of polarization development in a scattering region with
a torus shape. In the case of an optically thick and tall torus, diffusion
along the vertical direction tends to orient the electric field in the direction
perpendicular to the symmetry axis. In the case of an optically thin torus,
emergent light is polarized parallelly to the symmetry axis.
}
\label{torus_polarization}
\end{figure}

In Fig.~\ref{torus_polarization}, we illustrate the development of polarization 
in a torus model. In the tall and optically thick torus, photons tend to be polarized
in the direction perpendicular to the vertical axis as they diffuse along the vertical 
direction through a large number of scattering near the inner wall.
This behavior is similar to one considered in the optically thick case 
shown in Fig.~\ref{slab_polarization} and
explains the polarization flip in Fig.~\ref{spec_torus_flat}.
In the case of an optically thin torus, emergent radiation is polarized
in the direction parallel to the symmetry axis. This is also similar to
the polarization behavior found in the optically thin slab model.

\subsubsection{Effects of \ion{H}{I} Column Density and Covering Factor}
%
%\begin{figure}
%\includegraphics[scale=0.69]{tflux_torus_flat_bw.ps}
%\caption{Angular distribution of the total flux Rayleigh scattered
%from a torus. The incident radiation is a flat continuum and the vertical
%scale is logarithmic.
%}
%\label{tflux_torus_flat}
%\end{figure}
%
%

\begin{figure}
\includegraphics[scale=0.69]{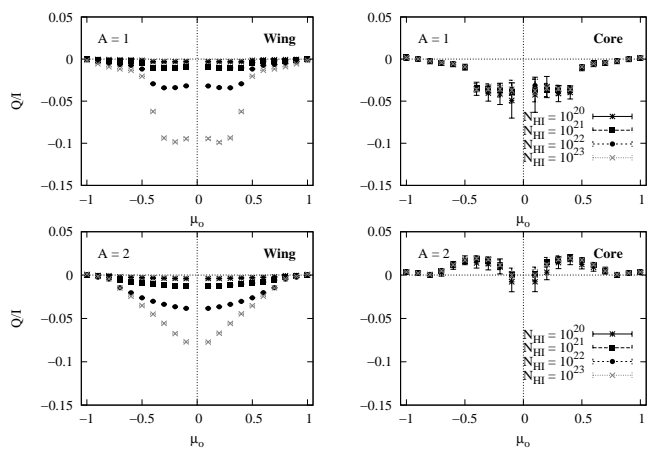}
\caption{Angular distribution of the polarization of Rayleigh scattered
from a torus. The left panels are for wing photons $\tau_s<10$
and the right panels are for core photons ($\tau_s>10$).
}
\label{tpol_torus_flat}
\end{figure}

In Fig.~\ref{tpol_torus_flat} we show
the angular distributions of the polarization of Rayleigh scattered
from tori with the two values of $A=1$ and $A=2$.
%The torus geometry is described
%by the ratio $A=H/R_i$ of the height and the inner radius, where
%the two values of $A=1$ and $A=2$ are considered in this work.
Four different values of $N_{HI}$ are investigated.
In the figure, we divide the incident radiation by the 
scattering optical depth $\tau_s$ in order to clarify the different
behaviors of wing photons and core ones. 

In the left panels we show the polarization of wing photons ($\tau_s < 10$),
all of which are polarized in the direction parallel to the symmetry axis.
In the top left panel for the case of $A=1$, we notice that maximum 
degree of polarization is not seen for grazingly emergent radiation 
but $\mu_o\simeq 0.2$. This is in contrast with the case of $A=2$
shown in the bottom left panel. 
In the case of $A=1$, the polarization behavior resembles that of the wing 
photons reflected in a slab illustrated in Fig.~\ref{tpol_slab_mock}. 
However, in the case of $A=2$ the polarization behavior resembles that of the transmitted component.

Totally different behaviors are noticed in the case of the transfer of
core photons ($\tau_s > 10$). In the top right panel for $A=1$ all the 
core photons are polarized in the direction parallel to the symmetry axis,
which is the same as their wing counterparts. The degree of polarization
is smaller due to the increased number of scatterings required before escape
for core photons. However, in the case of $A=2$ shown in the bottom
right panel, core photons are polarized in the perpendicular
direction. In this scattering geometry, an core photon needs to climb
up along the inner wall of the cylindrical shell by repeated bouncings 
before it makes the final exit of the scattering region. In the random walk
type process of moving up and down along the inner wall, the electric field
associated with the photon relaxes in the direction perpendicular to the
symmetry axis. This situation is analogous to the perpendicular 
polarization that develops in the very optically thick slab investigated by \citet{chandrasekhar60}.

\subsection{Mock Spectrum of AGNs in a Torus}

\begin{figure}
\includegraphics[scale=0.69]{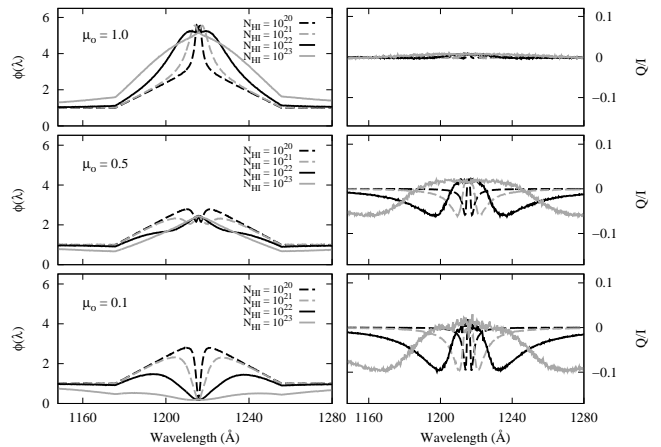}
\caption{Degree of polarization and flux of broad emission and flat continuum incidence source. In this torus model, the shape parameter $A$ is $2$.
}
\label{spec_torus_mock}
\end{figure}

\begin{figure}
\includegraphics[scale=0.69]{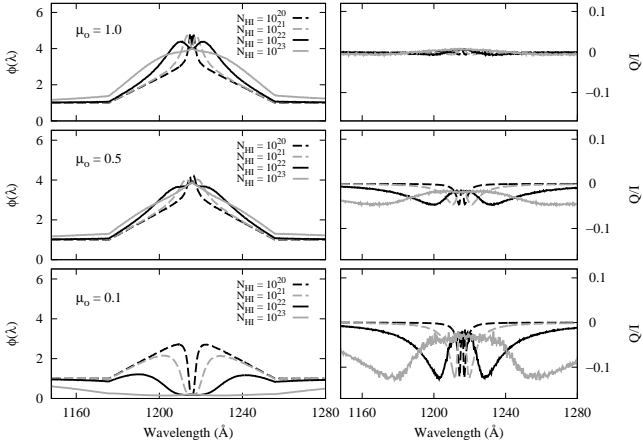}
\caption{This figure is composed by same scheme as in Fig.~\ref{spec_torus_mock} and $A = 1$. 
}
\label{spec_torus_mock_A1}
\end{figure}

In Fig.~\ref{spec_torus_mock}, we show our Monte Carlo simulated profiles 
and polarization of emergent radiation that is Rayleigh scattered
in a torus region. The incident radiation consists of a flat continuum
and a broad Ly$\alpha$ emission component with a triangular profile,
which was considered in section 3.2.3. 

The top panels show the data for radiation emergent nearly along
the symmetry axis, which is negligibly polarized. However, the profiles
are augmented by the Rayleigh scattered flux. For small $N_{HI}=10^{20}
{\rm\ cm^{-2}}$ Rayleigh scattering is limited to photons near line center
leading to a spiky central feature.

In the middle panels for emergent radiation with $\mu_o\sim 0.5$, 
the behavior of polarization flip is apparent. The center part
where the scattering optical thickness exceeds unity is polarized
in the perpendicular direction. The wing regions exhibit polarization
in the parallel direction due to dominant contribution from a
small number of scattering occurring in the $x-y$ plane.

The bottom panel shows the simulated data for grazingly emergent
radiation. The largest degree of polarization is obtained
near the shoulder region with moderate scattering optical
depth. In the far wing region where the scattering optical depth
is very small, the degree of polarization is small due to the contribution 
from the unpolarized incident
radiation that is not Rayleigh scattered. 

Fig.~\ref{spec_torus_mock_A1} shows our Monte Carlo result for a short torus, for which
the behavior is in high contrast with that illustrated in Fig.\ref{spec_torus_mock}.
In particular, in the middle panel, one can see that Rayleigh scattered radiation
near Ly$\alpha$ is polarized in the direction parallel to the symmetry axis and
no polarization flip appears. In the bottom panel for which the anisotropy is 
maximal, the degree of polarization is enhanced and particularly the shoulder
part exhibits the degree of polarization in excess of 10 percent.

%\begin{figure}
%\includegraphics[scale=0.69]{tflux_torus_mock_bw.ps}
%\caption{Total flux from a torus and continuum + emission.
%}
%\label{tflux_torus_mock}
%\end{figure}

%\begin{figure}
%\includegraphics[scale=0.69]{tpol_torus_mock_bw.ps}
%\caption{Total polarization from a torus and continuum + emission.
%}
%\label{tpol_torus_mock}
%\end{figure}

\section{Discussion and Summary}

In this work, we compute the flux and polarization of the far UV radiation around Ly$\alpha$ 
Rayleigh scattered in a slab and a torus in order to develop a new spectropolarimetric
tool to investigate the unification models of AGNs. In the slab geometry, 
the transmitted radiation is weakly polarized with the profile exhibiting a central dip in the flux 
whereas the Rayleigh reflected radiation is significantly 
polarized with an enhanced line center. 

In the torus geometry we find an interesting 
behavior of polarization flip dependent on the shape of the torus. When the torus 
is tall with the shape parameter $A=H/R_i=2$, the Rayleigh thick part near the
line center is polarized in the direction perpendicular to the symmetry axis whereas
the wing parts are polarized in the parallel direction. However in the case
of a short torus model with $A=1$, no polarization flip is observed and the emergent
radiation is polarized in the direction parallel to the symmetry axis.

It has been proposed that hard X-ray background radiation is significantly
contributed by type 2 AGNs \citep[e,g.,][]{comastri95}. However,
the population of type 2 AGNs is only poorly constrained \citep{gilli07}. 
According to AGN unification models,
type 2 AGN population is closely related with the geometrical covering factor
of the molecular torus. In view of this, detection of polarized Ly$\alpha$
exhibiting a polarization flip in the wing parts may indicate the presence
of a tall molecular torus implying a fairly large population of type 2 AGNs.

The models adopted in this work are highly ideal and 
the \ion{H}{I} distribution is more probably far from being axisymmetric.
In this case, spectropolarimetric
observations may reveal  Ly$\alpha$ polarized in different directions depending
on wavelength because the Rayleigh scattering optical thickness has no axisymmetry.
In this work we made no consideration of narrow Ly$\alpha$ emission component,
which may fill the central dip shown in the Rayleigh transmitted component and
introduce an additional polarized component.

The Rayleigh scattering cross section is given as a squared sum consisting of contributions
from each bound $np$ and free $n'p$ state. For a photon in the vicinity of Ly$\alpha$, 
the $2p$ state provides a dominant contribution.
A minor contribution from the remaining $p$ states affects
the final cross section in a complicated way so that 
the cross section is larger for radiation redward of Ly$\alpha$ than blueward \citep{lee13}.

More quantitatively, for red and blue wavelengths around Ly$\alpha$,
$\lambda_{\pm}=(1\pm\Delta V/c)\lambda_\alpha$
%-\lambda_\alpha=\pm(\Delta V/c)\lambda_\alpha$
corresponding to the Doppler velocity $\Delta V=5,000{\rm\ km\ s^{-1}}$,
the asymmetry amounts to
\begin{equation}
{\sigma(\lambda_+) \over \sigma(\lambda_-)}-1\simeq 0.07.
\end{equation}
The asymmetry amounting to 7 percent in scattering cross section will be more easily
detected in the polarized flux than in the total flux.
A small excess in the polarized flux redward of Ly$\alpha$ can be regarded as a convincing 
signature of Rayleigh scattering.

Lying at $+5,900\ {\rm km\ s^{-1}}$ redward of
Ly$\alpha$, the broad \ion{N}{V}$\lambda$1240 line is often severely blended
with the broad Ly$\alpha$ in many AGNs. This leads to the possibility 
that the \ion{N}{V} line photons can be 
Rayleigh scattered, which will also enhance the red part of Ly$\alpha$.
However, the red enhancement due to \ion{N}{V} is expected to peak at
$\Delta V=+5,900{\rm\ km\ s^{-1}}$, whereas the red enhancement due to
asymmetry in cross section is featureless.

%Recent spectropolarimetric observations of
%several quasars by the Hubble Space Telescope showed a steep increase of 
%the degree of polarization blueward of
%the Lyman edge (Koratkar et al. 1998). In particular, in the quasar 
One notable example is provided by \cite{koratkar95}, who used the {\it Hubble Space Telescope}
to perform spectropolarimetry of the quasar 
PG~1630+377. They reported that Ly$\alpha$ in this object is significantly polarized with the degree of polarization
7 per cent after correction 
for unpolarized continuum, whereas no evidence of strong polarization was 
found in other lines. This is an interesting case because only Ly$\alpha$ can be
significantly polarized when the main scattering mechanism is Rayleigh. However,
their spectral resolution was too insufficient to reveal any polarization structures, which hinders
one from drawing a definite conclusion.
Future spectropolarimetric observations will shed much light on the AGN unification model.

\section*{Acknowledgements}

We are grateful to the anonymous referees for their useful comments, which
improved the presentation of this paper.
This research was supported by the Korea Astronomy and Space Science 
Institute under the R\&D program (Project No. 2015-1-320-18) supervised 
by the Ministry of Science, ICT and Future Planning.
This research also received partial support from the Basic Science Research Program through the National
Research Foundation (NRF-2014R1A1A2054887).
Y.Y. acknowledges the partial support from Basic Science Research Program
through the National Research Foundation of Korea (NRF) funded by the
Ministry of Science, ICT \& Future Planning (NRF-2016R1C1B2007782).

%%%%%%%%%%%%%%%%%%%% REFERENCES %%%%%%%%%%%%%%%%%%

% The best way to enter references is to use BibTeX:

%\bibliographystyle{mnras}
%\bibliography{example} % if your bibtex file is called example.bib

% Alternatively you could enter them by hand, like this:
% This method is tedious and prone to error if you have lots of references

%%%%%%%%%%%%%%%%%%%%%%%%%%%%%%%%%%%%%%%%%%%%%%%%%%

%%%%%%%%%%%%%%%%%%%%%%%%%%%%%%%%%%%%%%%%%%%%%%%%%%

% Don't change these lines
\bsp	% typesetting comment
\label{lastpage}
\end{document}